\documentclass[graybox]{svmult}

\usepackage{type1cm}        
\usepackage{makeidx}         
\usepackage{graphicx}        
\usepackage{multicol}        
\usepackage[bottom]{footmisc}

\usepackage{newtxtext}       %
\usepackage[varvw]{newtxmath}       

\makeindex             

\usepackage{color}
\usepackage{amsmath}
\usepackage{bm}
\usepackage{physics}

\begin{document}

\title*{Novel understanding of Cosmological Phenomena using Fast Radio Bursts}
\titlerunning{Fast radio bursts to probe cosmological phenomena} 
\authorrunning{Kalita, Bhatporia \& Weltman} 

\author{Surajit Kalita\orcidID{0000-0002-3818-6037} \and \\ Shruti Bhatporia\orcidID{0000-0003-2821-4927} \and \\ Amanda Weltman\orcidID{0000-0002-5974-4114}}
\institute{Surajit Kalita \at High Energy Physics, Cosmology and Astrophysics Theory (HEPCAT) Group, \\ Department of Mathematics and Applied Mathematics, University of Cape Town, \\ Cape Town 7700, South Africa, \email{surajit.kalita@uct.ac.za}
\and Shruti Bhatporia \at High Energy Physics, Cosmology and Astrophysics Theory (HEPCAT) Group, \\ Department of Mathematics and Applied Mathematics, University of Cape Town, \\ Cape Town 7700, South Africa \email{shrutibhatporia@gmail.com}
\and Amanda Weltman \at High Energy Physics, Cosmology and Astrophysics Theory (HEPCAT) Group, \\ Department of Mathematics and Applied Mathematics, University of Cape Town, \\ Cape Town 7700, South Africa \email{amanda.weltman@uct.ac.za}
}
%
%
\maketitle

\abstract{Fast radio bursts (FRBs) offer unique probes of diverse cosmological phenomena due to their characteristic properties, including short duration timescale and high dispersion measure. This study investigates two distinct theoretical frameworks: the Gertsenshtein-Zel'dovich (GZ) mechanism for ultra-high-frequency gravitational waves (GWs) and fraction of dark matter in primordial mass black holes. We explore the hypothesis that ultra-high-frequency GWs could be responsible for FRB generation. Consequently, the detection of continuous GWs signal from the vicinity of an FRB by current or future detectors would disfavour merger-based FRB formation models and lend significant credence to the GZ theory, which postulates the existence of high-frequency GWs. Moreover, we examine the effects of modified gravity on the gravitational lensing of FRBs and thereby put constraints on the fraction of primordial mass black holes made up of dark matter. Our analysis suggests that modified gravity introduces a screening effect on lensing, analogous to the scattering effect by plasma on light rays. We further discuss the expected detection rates of FRBs as well as lensed FRBs with upcoming radio telescopes, primarily HIRAX.}


\section{Introduction}

Fast radio bursts~(FRBs), characterized by millisecond radio transients, exhibit high fluxes on the order of hundreds of $\rm Jy$. Despite their recent discovery, FRBs have garnered significant scientific interest due to their unique properties and potential to unveil new astrophysical processes. Extragalactic distances for most FRBs are suggested by their high dispersion measures~(DMs) and isotropic sky distribution, favoring a cosmological origin. However, pinpointing their exact host environments is challenging. Associations with dwarf galaxies have been observed for some FRBs, while others might originate from regions with strong magnetic fields, such as those surrounding magnetars or neutron stars~(NSs). The recent discovery of a repeating FRB~(FRB\,20201124A)~\cite{2022NatCo..13.4382W} within a young stellar cluster suggests a possible connection to massive star formation regions.

The true origin of all FRBs remains largely enigmatic although the Galactic FRB\,20200428 confirms an association with the magnetar SGR\,1935+2154~\cite{2020PASP..132c4202B,2020Natur.587...59B,2020Natur.587...54C}. Their immense luminosity suggests a cataclysmic event; however, the lack of a confirmed electromagnetic~(EM) counterpart hinders definitive progenitor identification. Proposed candidates include giant magnetar flares, collapsars during core-collapse supernovae, or even processes involving NS mergers~\cite{2019PhR...821....1P}. Notably, all these mechanisms involve compact objects like white dwarfs~(WDs), NSs, or black holes~(BHs). Therefore, multi-wavelength observations or detections of gravitational waves (GWs) and neutrinos from FRBs are crucial to understanding their progenitor mechanism. Unveiling the FRB progenitor holds the potential to revolutionize our understanding of stellar evolution, interstellar medium (ISM) properties, and even fundamental physics concepts like magnetar behavior or the existence of exotic new objects.

FRBs hold immense potential as powerful tools for cosmological investigations and testing gravity theories. By studying DM, the delay experienced by different radio frequencies due to plasma along the propagation path, FRBs can map the total electron content along their path. This information, combined with redshift measurements (when possible), allows for mapping the distribution of matter across vast cosmic distances. Additionally, studying FRB polarization and potential interactions with the magneto-ionic medium can provide clues about the local environment around their source. Ultimately, FRBs offer a unique opportunity to map the large-scale structure of the Universe, constrain cosmological parameters, and potentially shed light on fundamental physics at the extreme environments from which they originate.

In this article, we aim to explore the applications of FRBs in cosmology, particularly with gravitational lensing and high-frequency gravitational waves. In Sec.~\ref{Sec2}, we discuss how GW astronomy would help in understanding the progenitor mechanism of FRBs. Moreover, in Sec.~\ref{Sec3}, using lensing in FRBs, we obtain the constraints on the fraction of dark matter in primordial mass black holes in modified gravity theory. Furthermore, in Sec.~\ref{Sec4}, we estimate the total FRB and lensed FRB rates detectable with the upcoming Hydrogen Intensity and Real-time Analysis eXperiment (HIRAX) telescope in South Africa. Finally, we put our concluding remarks in Sec.~\ref{Sec5}.

\section{Gravitational waves to distinguish progenitor theories of FRBs}\label{Sec2}

As mentioned in the previous section, several astrophysical mechanisms have been proposed to explain the origins of FRBs. These mechanisms can be broadly categorized into two regimes: those involving the coalescence of compact objects and those involving isolated compact objects. In the merger scenario, a powerful magnetohydrodynamic shock wave is generated during the collision. Electrons accelerated within this shock dissipate a significant amount of their energy within the surrounding magnetosphere, potentially leading to the production of an FRB. Notably, the final merger remnant, likely a black hole, would no longer be expected to emit significant gravitational radiation. In contrast, FRBs arising from isolated compact objects would leave the object in its original location. Future generation GW detectors with enhanced sensitivity may be capable of directly detecting these objects, provided they retain rotation-powered emissions similar to those of pulsars.

\begin{figure}
    \centering
    \includegraphics[scale=0.4]{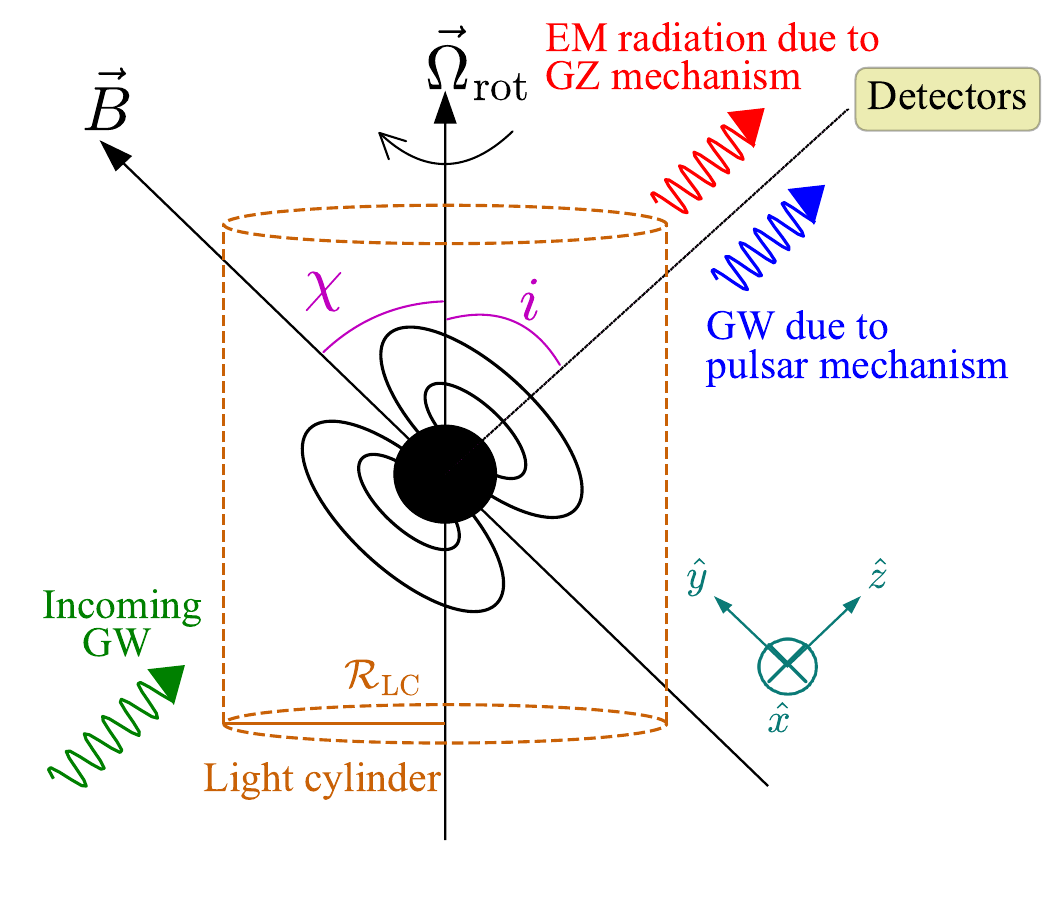}
    \caption{Schematic diagram of GZ mechanism showing green GWs get converted to red EM waves due to its interaction with the pulsar magnetosphere. The blue GWs are emitted due to the pulsar mechanism.}
    \label{Fig:GZ}
\end{figure}

We now focus on a specific scenario for the generation of FRBs from isolated compact objects: the Gertsenshtein-Zel'dovich (GZ) effect. It proposes a partial conversion of incoming GWs into EM waves upon interaction with the magnetic field of the pulsar magnetosphere. If the resulting EM radiation possesses characteristics consistent with observed FRBs, the GZ mechanism becomes a viable candidate for FRB production. Fig.~\ref{Fig:GZ} illustrates a schematic representation of this effect. Let us assume the polarization modes over time $t$ of infall GWs propagating in $z$-direction are given by
\begin{align}
    h_+ = A_+ e^{i\left(k_g z - \Omega_g t\right)} \quad \text{and} \quad
    h_\times = i A_\times e^{i\left(k_g z - \Omega_g t\right)},
\end{align}
where $\Omega_g$ is the GW frequency, $k_g$ is its wave number, $A_+$ and $A_\times$ are their amplitudes. Assuming $A_+=A_\times$ and the time-varying magnetic field of the pulsar in $y$-direction as $B_y(t) = B^{(0)}_y + \delta B_y \sin(\Omega_\mathrm{rot} t)$ with $\Omega_\mathrm{rot}$ being the rotation frequency of the pulsar, outgoing EM waves can be represented the following wave equations~\cite{2024MNRAS.527.4378K}
\begin{align}
    \frac{1}{c^2}\pdv[2]{\tilde{E}_x}{t} -\partial_z^2{\tilde{E}_x} = f_E(z,t)  \quad \text{and} \quad
    \frac{1}{c^2}\pdv[2]{\tilde{B}_y}{t} -\partial_z^2{\tilde{B}_y} = f_B(z,t),
\end{align}
where $f_E, f_B \sim e^{i\left(k_g z - \Omega_+ t\right)}$ and $\Omega_{\pm} = \Omega_g \pm \Omega_\mathrm{rot}$. Note that as $\Omega_\pm \sim \rm MHz-GHz$ (due to observed frequency of FRBs) and $\Omega_\mathrm{rot} \sim \rm Hz-kHz$, we must have $\Omega_g \sim \rm MHz-GHz$. This implies that the infall gravitational radiation must be cosmological in nature. For a detailed discussion on different physical mechanisms that emit GWs at frequencies $\rm MHz-GHz$, one may refer \cite{2021LRR....24....4A}.

\begin{figure}
    \centering
    \includegraphics[scale=0.35]{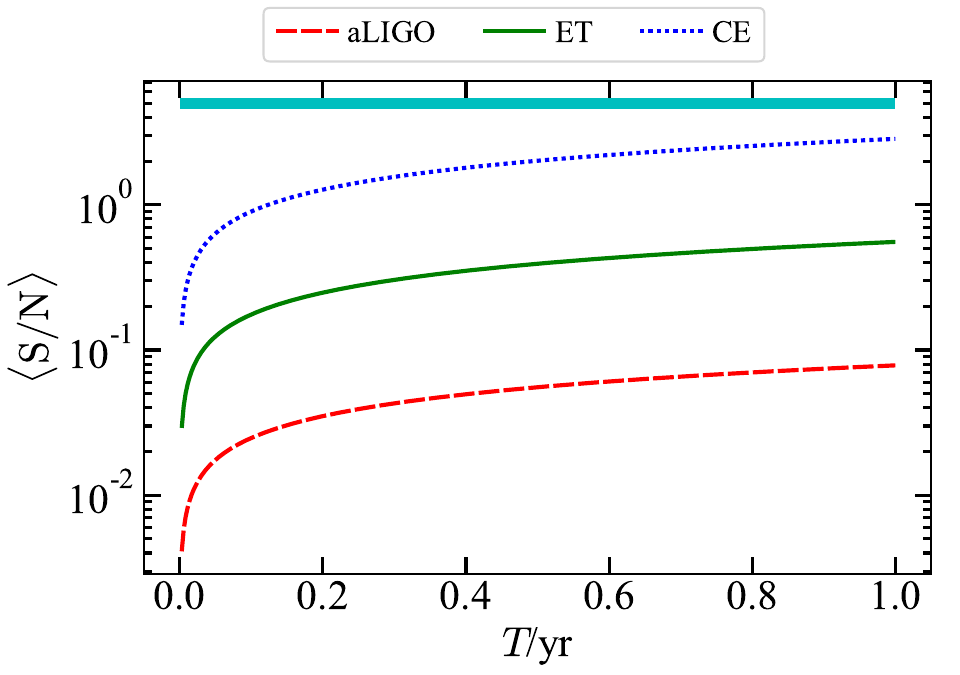}
    \includegraphics[scale=0.35]{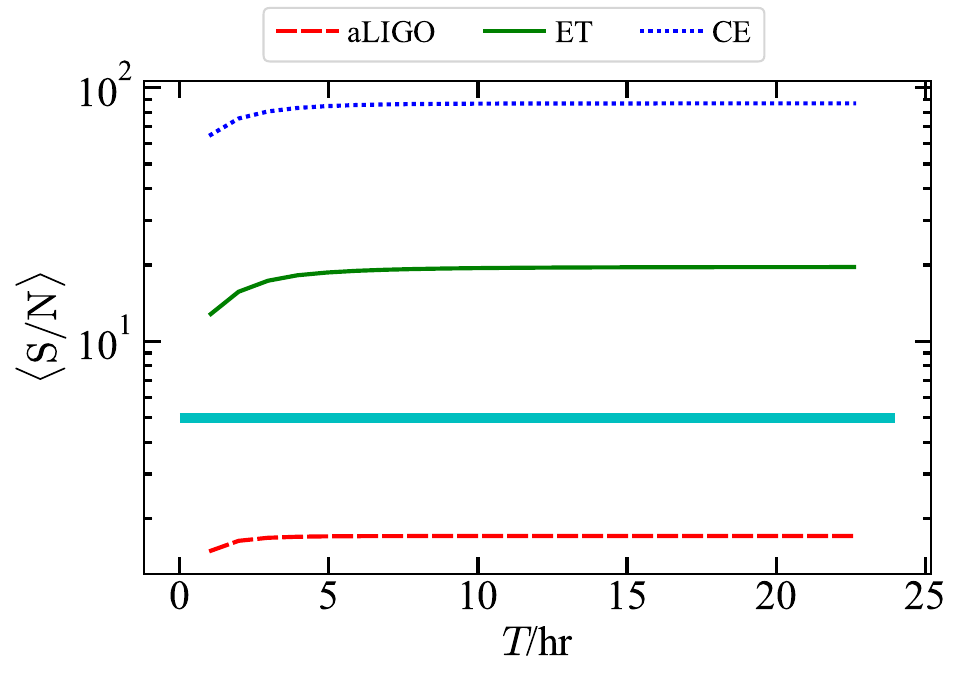}
    \caption{Cumulative signal-to-noise ratio (S/N) over time for different GW detectors. The left figure shows a typical case of NS whereas the right figure is for a magnetar. The thick cyan line represents the threshold S/N above which the signal can be detected.}
    \label{Fig:GW}
\end{figure}

Let us now illustrate the concept with FRB\,20180817, an FRB observed at a frequency of 501.1\,MHz with pulse width $\delta=0.01769\rm\,s$ and peak flux = 2.4\,Jy. This translates to a light cylinder radius $\mathcal{R}_\mathrm{LC} = \delta c/2 = 2.65\times10^{8}\rm\,cm$ and thereby $\Omega_\mathrm{rot} = c/\mathcal{R}_\mathrm{LC} = 113.1\rm\,rad\,s^{-1}$. Now, the peak flux is theoretically given by
\begin{equation}\label{Eq: peak flux}
    S_z = \frac{A_+^2 \abs{B^{(0)}_y}^2 c}{128\pi}\left[\sqrt{\frac{24 c^2 \nu_g^2 \alpha_\mathrm{tot}}{\pi G \abs{B^{(0)}_y}^2}-51} - \frac{6 c^2 \nu_g \nu_\mathrm{rot} \alpha_\mathrm{tot}}{\pi G \abs{B^{(0)}_y}^2}-1\right],
\end{equation}
where $c$ is the speed of light, $G$ is Newton's gravitational constant, $\nu_{g,\mathrm{rot}} = \Omega_{g,\mathrm{rot}}/2\pi$, and $\alpha_\mathrm{tot}$ is the total amount of energy converted from GWs to EM waves. Substituting the known values results in two unknowns: $A_+$ and $\abs{B^{(0)}_y}$. Assuming $A_+=10^{-24}$, we obtain $\abs{B^{(0)}_y} \approx 3\times10^{10}\rm\,G$ and if $A_+=10^{-29}$, we obtain $\abs{B^{(0)}_y} \approx 3\times10^{15}\rm\,G$. These values suggest the compact object could either be a NS or a magnetar. To definitively distinguish between the two, we require GW detection from the FRB's location. A detailed analysis of GW detection for such scenarios can be found in~\cite{2023MNRAS.520.3742K}. Fig.~\ref{Fig:GW} illustrates the possibility of detecting GWs from NS- and magnetar-based scenarios. For magnetars, detectors like the Cosmic Explorer and the Einstein Telescope could potentially detect the signal immediately. However, for NS, none of the currently proposed detectors is sensitive enough. Therefore, GW observations from the FRB's location hold the key to differentiating between the merger and isolated progenitor mechanisms for FRBs.

\section{FRBs to constrain dark matter fraction in primordial mass black holes}\label{Sec3}

Beyond their origin, FRBs serve as powerful cosmological probes. Their high DM information can shed light on the properties of the intergalactic medium~(IGM) and the large-scale structure of the cosmos. Additionally, studying lensed FRBs, where the light from an FRB is bent by the gravity of a massive object along its path, can provide unique insights into dark matter distribution and the nature of gravity itself. In strong lensing events, a single FRB source can be magnified and split into multiple images. Considering spherical polar coordinates $(t,r,\theta,\phi)$, let us now assume the following spherically symmetric background metric 
\begin{align}
    \dd{s}^2 = \left(1-\frac{2GM_\mathrm{L}}{c^2r}+\Psi r^2\right)c^2\dd{t}^2 - \frac{1}{1-\frac{2GM_\mathrm{L}}{c^2r}+\Psi r^2}\dd{r}^2 -r^2 \dd\theta^2-r^2\sin^2\theta \dd\phi^2,
\end{align}
where $M_\mathrm{L}$ is the lensed mass and $\Psi$ is a parameter characterizing modified gravity. We consider this particular metric as it is a solution within the framework of a general scalar-tensor gravity theory. Thereby, employing Shapiro delay on the lensed images, we can obtain the differential time delay between the two images, which is given by~\cite{2023JCAP...11..059K}
\begin{align}
    \Delta t = \left(1+z_\mathrm{L}\right) \left[\frac{1}{c\sqrt{\Psi}}\tan^{-1} \left(\sqrt{\Psi}\frac{4GM_\mathrm{L}}{c^2} \frac{y}{2}\sqrt{y^2+4}\right) + \frac{4GM_\mathrm{L}}{c^3} \ln\left(\frac{\sqrt{y^2+4} + y}{\sqrt{y^2+4} - y}\right)\right],
\end{align}
where $z_\mathrm{L}$ is the redshift of the lens plane, $y = r/\theta_\mathrm{E}\left(D_\mathrm{L}+D_\mathrm{LS}\right)$ with $\theta_\mathrm{E}$ being the Einstein radius, and $D_\mathrm{L}$ and $D_\mathrm{LS}$ correspond to the angular diameter distances from the observer to the lensing plane and from the lensing plane to the source, respectively.

This analysis utilizes the Canadian Hydrogen Intensity Mapping Experiment (CHIME) dataset\footnote{\url{https://www.chime-frb.ca/catalog}}, encompassing 636 FRBs. The total DM can be decomposed into individual contributions from various sources as follows:
\begin{align}
    \mathrm{DM} = \mathrm{DM}_\mathrm{MW} + \mathrm{DM}_\mathrm{Halo} + \mathrm{DM}_\mathrm{IGM}(z_\mathrm{s}) + \frac{\mathrm{DM}_\mathrm{Host}}{1+z_\mathrm{s}},
\end{align}
where $\mathrm{DM}_\mathrm{MW}$, $\mathrm{DM}_\mathrm{Halo}$, $\mathrm{DM}_\mathrm{IGM}$, and $\mathrm{DM}_\mathrm{Host}$ are DM contributions from our Milky way galaxy, its circumgalactic halo, IGM, and host galaxy, respectively with $z_\mathrm{s}$ being the source redshift. Assuming standard $\Lambda$CDM cosmology, the mean DM contribution from IGM is given by~\cite{2020Natur.581..391M}
\begin{align}
    \langle \mathrm{DM}_\mathrm{IGM}\rangle = \frac{3cH_0\Omega_\mathrm{b}}{8\pi G m_\mathrm{p}} \int_{0}^{z_\mathrm{s}} \frac{f_\mathrm{IGM}(z)\chi(z)(1 + z)}{\sqrt{\Omega_\mathrm{m} \left(1+z\right)^3 + \Omega_\Lambda}} \dd{z},
\end{align}
where $H_0$ is the Hubble constant, $\Omega_\mathrm{b}$ is the baryonic matter density, $m_\mathrm{p}$ is the proton mass, $\chi(z)$ is the ionization fraction along the line of sight, and $f_\text{IGM}$ is the baryon mass fraction in the IGM. Employing fixed value of $\langle\mathrm{DM}_\mathrm{Host}\rangle = 117\rm\,pc\,cm^{-3}$ and calculating $\mathrm{DM}_\mathrm{MW}$, $\mathrm{DM}_\mathrm{Halo}$ using the NE2001 model of Galactic distribution of free electrons,  we determine $z_\mathrm{s}$ for all 636 FRBs in the dataset. Note that minor adjustments to $\langle\mathrm{DM}_\mathrm{Host}\rangle$ are expected to have a negligible impact on our overall results.

\begin{figure}
    \centering
    \includegraphics[scale=0.5]{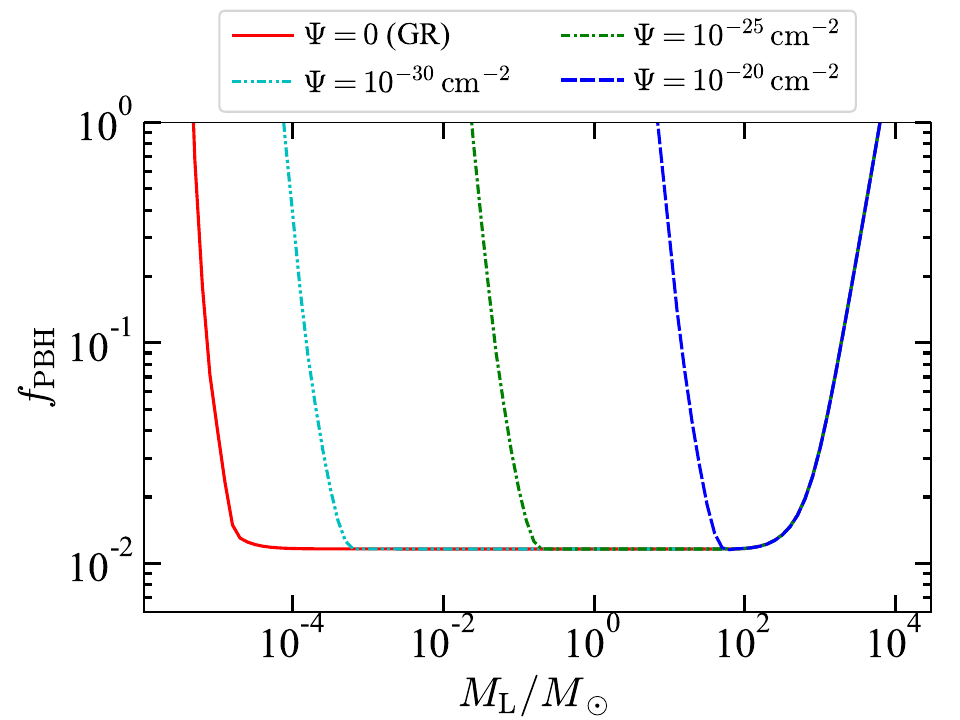}
    \caption{Constraints on fraction of dark matter in primordial mass black holes for different values of modified gravity parameter $\Psi$.}
    \label{Fig:f_bound}
\end{figure}

Considering the possibility of the lensed object being a primordial black hole~(PBH), we now define the optical depth of an FRB source~\cite{2016PhRvL.117i1301M}
\begin{align}
    \tau(M_\mathrm{L},z_\mathrm{s}) &= \frac{3}{2c}f_\mathrm{PBH} \Omega_\mathrm{c}\int_0^{z_\mathrm{S}} \dd{z_\mathrm{L}} \frac{H_0}{\sqrt{\Omega_\mathrm{m} \left(1+z_\mathrm{L}\right)^3 + \Omega_\Lambda}} \frac{D_\mathrm{L} D_\mathrm{LS}}{D_\mathrm{S}} \left(1+z_\mathrm{L}\right)^2 \nonumber \\& \times \left[y_\mathrm{max}^2(\mu) - y_\mathrm{min}^2(M_\mathrm{L},z_\mathrm{L})\right],
\end{align}
where $f_\mathrm{PBH}$ is the fraction of dark matter made up of PBHs, $\Omega_\mathrm{c}$ is the current cold dark matter density, $D_\mathrm{S}$ is the angular diameter distance to the source, $y_\mathrm{min}$ and $y_\mathrm{max}$ are respectively the minimum and maximum impact parameters. Taking into account all FRBs, we define the following integrated optical depth
\begin{align}
    \bar{\tau} = \frac{1}{N_\mathrm{FRB}}\sum_{i=1}^{N_\mathrm{FRB}} \tau (M_\mathrm{L},z_{\mathrm{s},i}).
\end{align}
As the number of lensed FRBs is expected to be quite small in comparison to the total number of FRBs, we can safely use the Poisson statistics as follows:
\begin{align}
    N_\mathrm{lensed, FRB} = \left(1-e^{-\bar{\tau}}\right) N_\mathrm{FRB}.
\end{align}
Since no lensed FRB has been confirmed so far, the above relation implies
\begin{align}
    f_\mathrm{PBH}<\frac{1}{\tau_1}\ln\left(\frac{N_\mathrm{FRB}}{N_\mathrm{FRB}-1}\right),
\end{align}
where $\bar{\tau}= f_\mathrm{PBH} \tau_1$. Fig.~\ref{Fig:f_bound} presents constraints on $f_\mathrm{PBH}$ across a range of modified gravity parameter values within the standard $\Lambda$CDM cosmological framework. Notably, the bound on $f_\mathrm{PBH}$ exhibits a significant dependence on the value of the parameter $\Psi$. The minimum detectable time delay of the observing telescope, approximately $10^{-9}\rm\,s$ for CHIME~\cite{2022PhRvD.106d3017L}, defines the left-hand cutoff for each curve. Conversely, the right-hand cutoff is determined by the maximum achievable magnification ratio of the lensed image pair, which essentially reflects the detectability of the fainter image alongside the brighter one. Interestingly, our results demonstrate similarity to those observed for plasma lensing~\cite{2022PhRvD.106d3017L}. It suggests that modified gravity, in this context, mimics the behavior of a scattering screen positioned along the path of the light ray. It is noteworthy that current experiments can, at best, place upper limits on $f_\mathrm{PBH}$ parameter, but do not yet provide definitive evidence for PBHs.

\section{New generation telescope: HIRAX}\label{Sec4}

\begin{figure}
    \centering
    \includegraphics[scale=0.5]{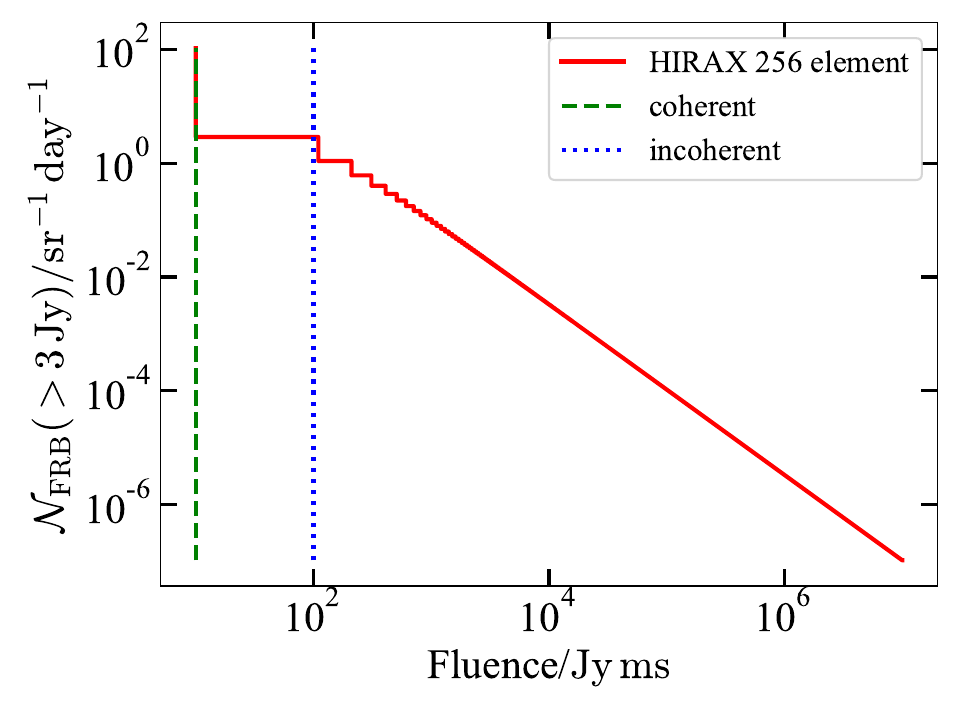}
    \caption{Expected FRB rates with HIRAX 256 elements. We show flux cutoff/telescope sensitivity for incoherent and coherent beam sets of FRBs. Even though computationally expansive, coherent beam configuration provides higher sensitivity of telescope.}
    \label{fig:hirax_dispersion_smearing_rate} 
\end{figure}

The quest to unravel the mysteries of FRBs is ongoing. In this study, we investigate the detection prospects of FRBs manifested by gravitational lensing with an under-construction HIRAX telescope in South Africa\footnote{\url{https://hirax.ukzn.ac.za}}. The effective detection rate of any object can be extrapolated observationally from the telescope's field of view~(FoV) and several detected sources for given flux/fluence limits~\cite{2015MNRAS.447.2852K}. Asssuming similarity between FRB and gamma-ray burst~(GRB) detection rates, the number of GRBs detectable by a given instrument above its sensitivity flux limit $S$ is given by~\cite{2016A&A...587A..40P}
\begin{equation}
    N(>S) = \frac{\Omega T}{4\pi} \int_{0}^{z(L_\mathrm{max},S)} \int_{L_\mathrm{min}(z,S)}^{L_\mathrm{max}} \Phi(L,z) \frac{\Psi(z)}{1+z} \dv{V}{z} \dd{L} \dd{z},
    \label{eq:gamma_ray_burst_rate}
\end{equation}
where $\Omega$ and $T$ are respectively system FoV and Temperature, $\Psi(z)$ is luminosity function, $\dv*{V}{z}$ is the differential comoving volume, and $\Phi(L,z)$ is GRB formation rate with $L$ being its luminosity and $z$ its redshift. This equation can be modified in FRB astronomy as~\cite{2023MNRAS.521.4024C}
\begin{equation}
    R_k = R_l \frac{\Omega_k}{\Omega_l} \left(\frac{\mathrm{SEFD}_l}{\mathrm{SEFD}_k}\sqrt{\frac{B_k n_{p,k}}{B_l n_{p,l}}} \right)^\alpha.
    \label{eq:frb_rate_survey}
\end{equation}
Here the rates $R$ are divided between two frequency bands and $\Omega$s are FoVs for those bands, $\mathrm{SEFD}$ is the corresponding system equivalent flux density, $B$ is bandwidth and $n$ is a number of polarizations. For current estimations, the source-count parameter $\alpha = 3R_l/2$, which is a known quantity, and $R_k$ is derived from Eq.~\eqref{eq:frb_rate_survey}. HIRAX is predicted to detect a few hundred FRBs per unit solid angle per day. In this regard, it has been proposed to develop a pipeline capable of detecting FRBs at various DMs in real-time. The following equation gives an estimation of FRB detection rate above an observed fluence $F_\mathrm{obs}$~\cite{2016ApJ...830...75V}
\begin{equation}
    \mathcal{N}(>F_\mathrm{obs}) = \frac{1.2\times10^4}{4\pi}\left(\frac{F_\mathrm{obs}}{1.8 \rm\,Jy\, ms}\right)^{-\alpha} \rm\,sr^{-1}\, day^{-1}.
    \label{eq:frb_rate_hirax}
\end{equation}
Fig.~\ref{fig:hirax_dispersion_smearing_rate} shows the FRB detection rate with HIRAX when the known rate is taken from the Parkes survey and all the calculations are done with uniform FRB source distribution with Euclidean sky.

The wider beamwidth of HIRAX facilitates a survey of large areas of the southern hemisphere. Current estimates predict HIRAX will detect on average several bursts daily utilizing its 256 elements. We analyze the likelihood of detecting lensed FRBs across a redshift range using the HIRAX/FRB search. We calculate the number of lensed FRBs expected at various redshifts considering different lensing sources throughout the year\footnote{Our results are based on the publicly available code \url{https://github.com/liamconnor/frb-grav-lensing}}. Our calculations incorporate microlensing~\cite{2023MNRAS.521.4024C} from sources like massive compact halo objects (MACHOs), PBHs, stars, free-floating planets, along with lensing by heavier objects like intermediate-mass black holes, dark matter halos, galaxy lensing at extended distances, and galaxy clusters. We assume a uniform distribution of both lensing sources and FRBs across the celestial sphere. Based on this, we predict roughly one lensing event per year detectable by HIRAX. We also incorporate results from CHIME and The Canadian Hydrogen Observatory and Radio-transient Detector~(CHORD) in this analysis~\cite{2019clrp.2020...28V}. 

\begin{figure}
    \centering
    \includegraphics[scale=0.5]{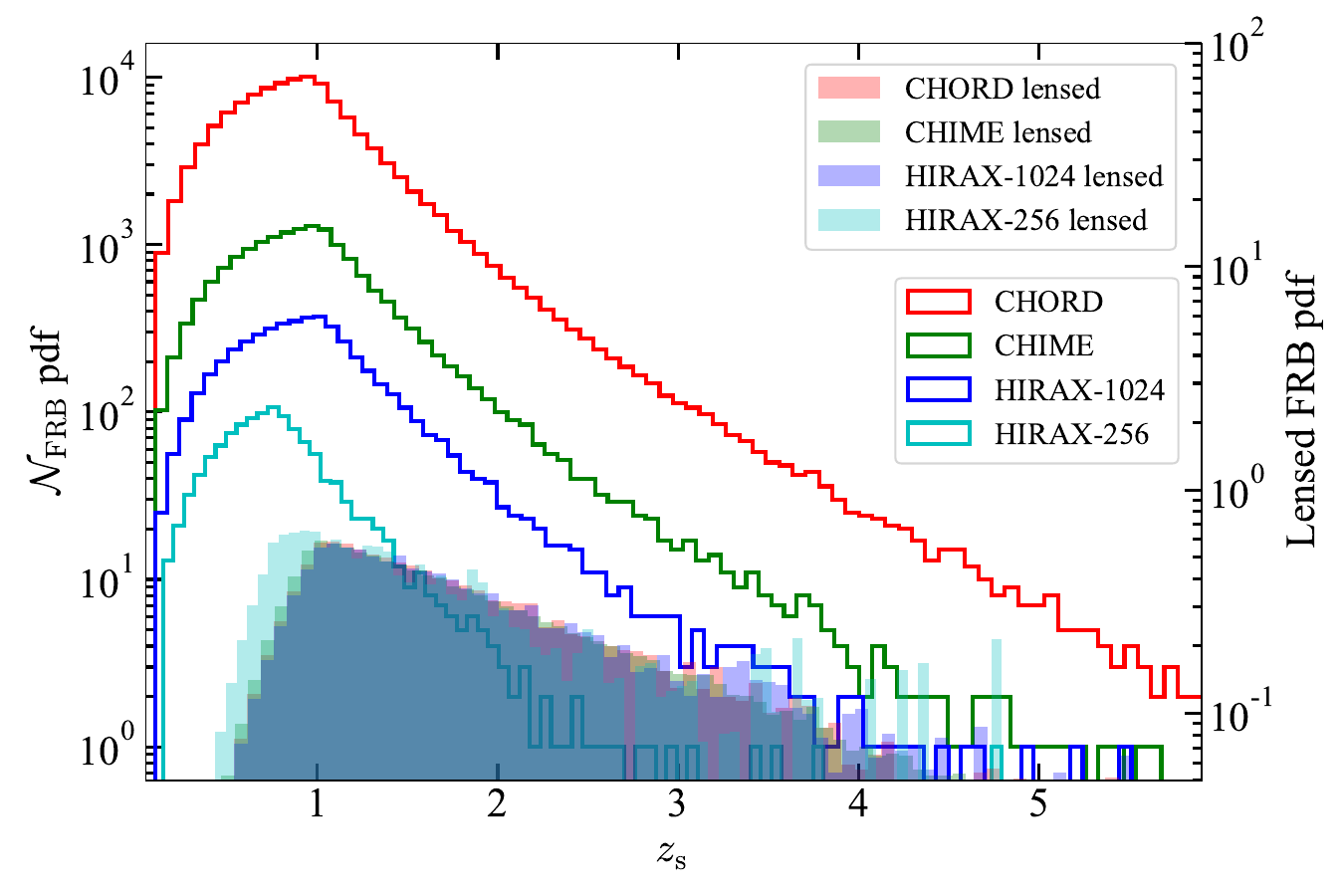}
    \caption{Distribution of FRBs over redshifts. We perform a comparative analysis of three radio telescopes: CHIME, its upgrade CHORD, and HIRAX. The total number of FRBs detected by HIRAX, as well as the number of lensed FRBs, exhibits a peak around a redshift of 1.}
    \label{fig:hirax_frb_lens_CDF}
\end{figure}

All these telescopes have ability to probe distant FRBs at high redshifts as illustrated in Fig.~\ref{fig:hirax_frb_lens_CDF}. This figure represents the distribution of rate of FRBs detected by CHIME, CHORD, and HIRAX (solid curves). The shaded regions represent the normalized probability distribution function for all lensed FRBs across the redshift range. Consequently, future telescopes like HIRAX, specifically positioned in the southern hemisphere, offer significant potential for precise and sensitive FRB searches, along with predictions of lensed FRB occurrences.

\section{Conclusions}\label{Sec5}

This study investigates the application of FRBs for understanding various cosmological phenomena. We posit that high-frequency GWs of cosmological origin could be a potential FRB progenitor mechanism. In this regard, the GZ effect, where GWs convert to EM waves upon interaction with the magnetosphere of a pulsar, offers a viable explanation for FRB generation. This mechanism is particularly attractive as it can accommodate both repeating and apparently non-repeating FRBs. Furthermore, the future detection of GWs co-located with FRBs could differentiate between merger-based and isolated progenitor scenarios. Additionally, it could provide insights into the nature of the compact object, which is currently challenging with radio observations alone. The second part of this work explores how the lensing properties of FRBs can be utilized to constrain the fraction of PBHs composed of dark matter. We further demonstrate that modified gravity can mimic a scattering screen for light rays, analogous to plasma lensing.

Pinpointing the location of FRBs with high precision can help us understand their environments. Studying the host galaxies of FRBs can reveal information about the conditions favorable for their formation. In this context, several next-generation radio telescopes with enhanced sensitivity and resolution are under construction, including HIRAX, the Square Kilometre Array~(SKA), CHORD, Deep Synoptic Array~(DSA)-2000, and Bustling Universe Radio Survey Telescope in Taiwan~(BURSTT) are being built. These telescopes hold immense promise for acquiring a wealth of new data, potentially leading to significant breakthroughs in FRB research. Our calculations estimate the FRB detection rate for HIRAX, revealing a rate comparable to CHORD, an upgraded version of the existing CHIME telescope. Both HIRAX and CHORD offer the capability to probe higher redshifts compared to CHIME, enabling the detection of fainter and more distant FRBs. Considering the rarity of astronomical objects detected at cosmological distances, the anticipated increase in FRB detections, with a significant number becoming localized, opens exciting avenues for exploring new physics, such as underscoring the possibility of resolving the Hubble tension or placing stronger limits on the photon mass in the near future.

\acknowledgement{We gratefully acknowledge support from the University of Cape Town Vice Chancellor’s Future Leaders 2030 Awards programme which has generously funded this research and support from the South African Research Chairs Initiative of the Department of Science and Technology and the National Research Foundation.}

\bibliographystyle{spphys}
\bibliography{Bibliography}
\end{document}